\documentclass[twocolumn,floatfix,aps,prb,showpacs,showkeys,superscriptaddress]{revtex4-1}
\usepackage[utf8]{inputenc}
\usepackage{color}
\usepackage{verbatim}
\usepackage{amsmath}
\usepackage{amssymb}
\usepackage{graphicx}
\usepackage[unicode=true,
 bookmarks=true,bookmarksnumbered=false,bookmarksopen=false,
 breaklinks=false,pdfborder={0 0 1},backref=false,colorlinks=true]
 {hyperref}
\hypersetup{
 pdfborderstyle=,linkcolor=blue,citecolor=blue,urlcolor=blue}

\makeatletter
\usepackage{bm}
\usepackage{float}

\newcommand{\be}{\begin{equation}}
\newcommand{\ee}{\end{equation}}
\newcommand{\bea}{\begin{eqnarray}}
\newcommand{\eea}{\end{eqnarray}}

\makeatother

\begin{document}
\title{Realistic indirect spin-interactions between magnetic impurities on
a metallic Pb(110) surface}
\author{Alejandro Rébola}
\affiliation{Instituto de Física Rosario - CONICET, Bv. 27 de Febrero 210 bis,
2000 Rosario, Argentina}
\email{arebola@ifir-conicet.gov.ar}

\author{Alejandro M. Lobos}
\affiliation{Facultad de Ciencias Exactas y Naturales Universidad Nacional de Cuyo,
5500 Mendoza, Argentina}
\affiliation{Consejo Nacional de Investigaciones Científicas y Técnicas (CONICET),
Argentina}
\begin{abstract}
Motivated by recent experiments, here we study the indirect interactions
between magnetic impurities deposited on top of a clean Pb(110) surface
induced by the underlying conduction electrons. Our approach makes
use of \textit{ab initio} calculations to characterize the clean Pb(110)
surface and avoids self-consistency, a feature that greatly reduces
the computational cost. In combination with 2$^{\text{nd}}$ order
perturbation theory in the microscopic \textit{s-d} exchange parameter
$J_{K}$ between a magnetic adatom and the conduction electrons, we
are able to systematically derive the Ruderman-Kittel-Kasuya-Yosida
(RKKY), the Dzyaloshinskii-Moriya (DM) and the anisotropic tensor
interactions emerging at the Pb(110) surface between magnetic impurities.
The only adjustable parameter is $J_{K}$, which is fitted to reproduce
the experiments. Our results show important anisotropy effects arising
both from the rectangular geometry of the (110) unit cell, and from
the strong Rashba spin-orbit interaction due to the broken
inversion symmetry at the Pb(110) surface. In addition to Pb(110),
the characterization of the indirect spin interactions described here
could be extended to other realistic metallic surfaces for weakly-coupled
impurities, and would enable to fabricate atomic-size nanostructures
with engineered interactions and on-demand magnetic properties, anticipating
useful applications in nanotechnology. 
\end{abstract}
\maketitle

\section{Introduction}

\label{sec:intro}

The Rudermann-Kittel-Kasuya-Yosida (RKKY) exchange interaction is
an indirect magnetic coupling between localized magnetic moments,
mediated by the conduction electrons in a metallic substrate \cite{ruderman54,Kasuya56_Theory_of_metallic_FM_and_AFM,Yosida56_RKKY}.
This type of interaction plays a crucial role in systems displaying
giant magnetorresistance \cite{Baibich88_GMR_in_layered_magnetic_superlattices},
heavy-fermion magnetism and quantum criticality \cite{Doniach77,Hewson_1993,Loehneysen07_Heavy_fermions_and_magnetic_QPTs},
and in dilute magnetic semiconductors \cite{Meilikhov2007,Wang2008}.
More recently, it has also been observed in atomic-scale magnetic
systems fabricated with scanning tunneling microscopy (STM) techniques
\cite{Wahl07_Exchange_Interaction_between_Single_Magnetic_Adatoms,Meier08_Single_atom_magnetization,Zhou2010,Khajetoorians12_Atom_by_atom_engineering_of_nanomagnets,Khajetoorians16_Tailoring_chiral_interactions,Esat16_Chemically_driven_QPT_in_a_Kondo_molecule}%
\begin{comment}
, e.g., through the evolution of the Kondo temperature when the interatomic
distance between Co adatoms in dimer and trimer configurations on
a Cu(100) surface is varied\cite{Wahl07_Exchange_Interaction_between_Single_Magnetic_Adatoms}.
Later, Zhou et al. \cite{Zhou2010} studied the RKKY exchange directly
in dimers and trimers configurations of magnetic adatoms of cobalt
on a Pt(111) surface, by fitting the single-atom magnetization curves,
measured using spin-polarized STM, to first-principles calculations 
\end{comment}
. In these atomic-sized structures the RKKY interaction plays a major
role. For instance, it has been recently proposed as a key ingredient
in magnetic atomic chains deposited on conventional superconductors
with a strong Rashba spin-orbit coupling (SOC), systems predicted
to host Majorana-fermion quasiparticles (MQP) \cite{Nadj-Perdge13_Majorana_fermions_in_Shiba_chains,
Klinovaja13_TSC_and_Majorana_Fermions_in_RKKY_Systems,Braunecker13_Shiba_chain}.
These works have triggered a great amount of theoretical and experimental
research seeking to observe MQPs, which could be instrumental in the
fabrication of qubits for topological quantum computaters. In recent
experimental works involving atomic Fe chains ontop of clean Pb(111)
or Pb(110) surfaces, preliminar evidence of MQPs have been reported
%\cite{NadjPerge14_Observation_of_Majorana_fermions_in_Fe_chains,Pawlak15_Probing_Majorana_wavefunctions_in_Fe_chains}.
\cite{Nadj-Perge2014,Ruby2015,Pawlak2016,Feldman16_High_resolution_Majorana_Shiba_chain}.

Assuming an idealized isotropic free-electron conduction band, the
standard result for the RKKY interaction is $J_{\text{RKKY}}\left(\mathbf{r}\right)\sim\cos\left(k_{F}r\right)/r^{D}$,
where $k_{F}$ the Fermi momentum and $r=\left|\mathbf{r}\right|$
the distance between the magnetic impurities is obtained \cite{ruderman54,Yosida56_RKKY,Kasuya56_Theory_of_metallic_FM_and_AFM}.
However, the behavior of real adatom systems on metallic surfaces
is strikingly different, and departures from an ideally isotropic
interaction has been reported experimentally. For instance, one of
the most relevant results in the abovementioned Refs. \onlinecite{Wahl07_Exchange_Interaction_between_Single_Magnetic_Adatoms, Meier08_Single_atom_magnetization, Zhou2010, Esat16_Chemically_driven_QPT_in_a_Kondo_molecule}
is the anisotropic character of the RKKY interaction on surfaces.
Considering the growing interest in the fabrication of magnetic devices
with specific functionalities and potential applications in quantum
computing, spintronic and magnetic memories, a detailed characterization
of realistic magnetic interactions would be highly desirable. From
a more fundamental perspective, a realistic characterization of the
RKKY interaction on specific metallic surfaces could also be useful
to simulate, in a controlled manner, the physics of strongly-correlated
materials. For instance, using self-assembled metal-organic networks
deposited on clean metallic surfaces, a controlled study of the celebrated
Kondo lattice model, typically used to understand the exotic low-temperature
behavior of heavy-fermion materials, has become possible with STM
techniques \cite{Tsukahara10_Evolution_of_Kondo_resonance,Komeda14_Adsorbed_molecules_investigated_by_Kondo_resonance,Girovsky2017_Ferrimagnetic_order_in_self_assembled_Kondo_lattice}.

Among the many possible metallic surfaces typically studied with STM,
the surface of Pb has become an ideal platform to study the interplay
between superconductivity and atomic magnetism. The interest is two-fold:
1) Pb becomes a conventional $s$-wave superconductor at low temperatures,
with a standard phonon-mediated pairing mechanism. In addition, its
relative simplicity to grow in films by evaporation techniques makes
it a widely used superconducting material in the laboratory. 2) A large
Rashba SOC exists at the surface of Pb, a property that is known to
induce large Dzyaloshinskii-Moriya interactions. This property could
be exploited in order to engineer non-colinear chiral magnetic nanostructures,
such as skyrmions\cite{Khajetoorians16_Tailoring_chiral_interactions,Bouaziz2017}.
Both features could prove extremely useful in novel spintronic devices
\cite{Linder2014,Linder2015,Krupin2005,Chuang2014,Cahay2004}. In
previous works, perturbative approaches combined with numerical and/or
semi-analytical methods for realistic band-structure calculations
have been used for the calculation of the RKKY indirect-exchange interaction
between nuclear moments\cite{Frisken86,Oja89,Patnaik98,Harmon92}.
However, none of these works have focused on magnetic impurities on
Pb, where relativistic effects are unavoidable. On the other hand,
the calculation of the Dzyaloshinskii-Moriya interaction has been
tackled in previous works using highly idealized model Hamiltonians
\cite{Kim2013,Kundu2015,Kikuchi2016,Bouaziz2017}, which ignore the
real electronic structure. Therefore, there are no systematic studies
of the realistic magnetic interactions on the surface of Pb.

Motivated by the aforementioned experimental advances, in this article
we focus on the derivation of realistic magnetic interactions between
impurities ontop of a clean Pb surface. For concreteness, and in order
to make contact with Refs. \onlinecite{Nadj-Perge2014,Ruby2015, Pawlak2016, Feldman16_High_resolution_Majorana_Shiba_chain},
we have chosen the particular case of Pb(110) surface. However, we
stress that our method is also applicable to other systems. Using
a combination of an analytical approach, i.e., second-order perturbation
theory in the \textit{s-d} exchange interaction $J_{K}$, and density
functional theory (DFT) to obtain the full band structure of Pb(110)
including relativistic SOC effects, we systematically derive the RKKY,
the Dzyaloshinskii-Moriya (DM) and the anisotropic tensor interactions
between magnetic impurities. Our results show important anisotropy
effects arising both from the rectangular geometry of the (110) unit
cell, and from the strong Rashba SOC originated in the broken inversion
symmetry at the Pb(110) surface.

Since within our perturbative approach only the band structure of
the \emph{clean} Pb(110) surface is needed, the computational cost
can be significatively reduced. This represents one of the main advantages
of our method: the possibility to describe indirect magnetic interactions
realistically (i.e., without having to resort to any \textit{a priori}
model or approximation), combined with low computational cost as compared
to standard self-consistent methods, such as the Korringa-Kohn-Rostoker
(KKR) method. Due to the nature of the method, its applicability is
in principle limited to weakly-coupled adatom systems satisfying the
condition $\rho_\text{3D}J_{K}\ll1$, where $\rho_\text{3D}$ is the density
of conduction states per unit volume at the Fermi energy. Such limitation
is, nevertheless, not severe as there exist many examples of systems
fulfilling this condition {[}e.g., metal-organic complexes such as
MnPc molecules \cite{Franke_2011} or iron(II) porphyrin molecules
\cite{Heinrich_2015} deposited ontop of Pb(110), where the organic
ligand of the molecule tends to isolate the effective magnetic moment
from the surface, leading to a small effective coupling $J_{K}${]}.
The only adjustable parameter in our formalism is therefore the \textit{s-d}
exchange parameter $J_{K}$, which is fitted to reproduce experiments\cite{Franke_2011}.

The rest of the paper is organized as follows. In Sec. \ref{sec:theory}
we present the theoretical model and the derivation of the generic
RKKY, DM and tensor interactions directly from the conduction-electron
propagators. In Sec. \ref{sec:methods} we give details about the
technical aspects of the \textit{ab initio} calculations and about
the convergence of the RKKY interaction. In Sec. \ref{sec:results}
we present the results, specifically in Sec. \ref{subsec:band_structure_pb110}
we present our results for the band-structure of the clean Pb(110),
and in Sec. \ref{subsec:magnetic_interactions} we show our results
for the magnetic RKKY, DM and tensor interactions. Finally, in Sec.
\ref{sec:summary} we summarize the main results and present the conclusions.

\section{Theoretical Model and Derivation of the Effective Interactions}

\label{sec:theory}

The theoretical model describing two spin impurities on a Pb(110)
surface, located at sites $\mathbf{r}_{1}=\left(x_{1},y_{1},0\right)$
and $\mathbf{r}_{2}=\left(x_{2},y_{2},0\right)$ where $z=0$ is the
coordinate of the surface plane, is 
\begin{align}
H & =H_{0}+H_{K}\left(1\right)+H_{K}\left(2\right).\label{eq:H_total}
\end{align}
Here

\begin{align}
H_{0} & =\sum_{\mathbf{k},n}\epsilon_{\mathbf{k},n}^{\left(0\right)}c_{\mathbf{k},n}^{\dagger}c_{\mathbf{k},n},\label{eq:H_0}
\end{align}
is the unperturbed Hamiltonian describing the bands of clean Pb(110).
The quantum numbers $\mathbf{k},n$ are, respectively, the crystal
momentum parallel to the surface belonging to the first Brillouin
zone, and the spin-orbital band index, which results from a combination
of the spin and the azimuthal angular momentum (recall that in the
presence of Rashba and/or Dresselhaus SOC, $s$, the spin projection
along $z$ is no longer a good quantum number. In the absence of Rashba
SOC, the index $n$ splits into $s$ and the usual band index $\alpha$).
The operator $c_{\mathbf{k},n}$ annihilates a fermionic quasiparticle
in the conduction band, and obeys the usual anticommutation relation
$\left\{ c_{\mathbf{k},n},c_{\mathbf{k}^{\prime},n^{\prime}}^{\dagger}\right\} =\delta_{\mathbf{k},\mathbf{k}^{\prime}}\delta_{n,n^{\prime}}$.
Finally, $\epsilon_{\mathbf{k},n}^{\left(0\right)}$ is the dispersion
relation computed in the absence of the magnetic impurities.

The Kondo (or \textit{s-d} exchange) interaction between a magnetic
moment and the conduction-electron spin density at point $\mathbf{r}_{j}$
is \cite{Hewson_1993} 
\begin{align}
H_{K}\left(j\right) & =J_{K}\mathbf{S}_{j}.\mathbf{s}\left(\mathbf{r}_{j}\right)\\
 & =J_{K}\mathbf{S}_{j}.\sum_{s,s^{\prime}=\left\{ \uparrow,\downarrow\right\} }\Psi_{s}^{\dagger}\left(\mathbf{r}_{j}\right)\left(\frac{\hat{\boldsymbol{\sigma}}}{2}\right)_{ss^{\prime}}\Psi_{s^{\prime}}\left(\mathbf{r}_{j}\right),\label{eq:H_K_fermion}
\end{align}
where $\hat{\boldsymbol{\sigma}}=\left(\hat{\sigma}_{x},\hat{\sigma}_{y},\hat{\sigma}_{z}\right)$
is the vector of Pauli matrices, and 
\begin{align}
\Psi_{s}\left(\mathbf{r}_{j}\right) & =\sum_{\mathbf{k},n}\psi_{\mathbf{k},n}^{\left(s\right)}\left(\mathbf{r}_{j}\right)c_{\mathbf{k},n},\label{eq:psi_real_space}
\end{align}
is the field operator which annihilates an electron with spin projection
$s=\left\{ \uparrow,\downarrow\right\} $ along the $z$ axis at point
$\mathbf{r}_{j}$, and $\psi_{\mathbf{k},n}^{\left(s\right)}\left(\mathbf{r}\right)$
are the normalized Bloch wavefunctions computed via DFT (see Section \ref{sec:methods}). The field operator obeys the usual relations: 
\begin{align}
\sum_{\mathbf{k},n}\psi_{\mathbf{k},n}^{*\left(s\right)}\left(\mathbf{r}_{i}\right)\psi_{\mathbf{k},n}^{\left(s^{\prime}\right)}\left(\mathbf{r}_{j}\right) & =\delta\left(\mathbf{r}_{i}-\mathbf{r}_{j}\right)\delta_{s,s^{\prime}},\label{eq:completeness}\\
\left\{ \Psi_{s}\left(\mathbf{r}_{i}\right),\Psi_{s^{\prime}}^{\dagger}\left(\mathbf{r}_{j}\right)\right\}  & =\delta\left(\mathbf{r}_{i}-\mathbf{r}_{j}\right)\delta_{s,s^{\prime}}.\label{eq:orthogonality}
\end{align}

The idea now is to use knowledge of the \emph{realistic} band structure
of Pb(110), encoded in $\epsilon_{\mathbf{k},n}^{\left(0\right)}$
and $\psi_{\mathbf{k},n}^{\left(s\right)}\left(\mathbf{r}_{j}\right)$,
in order to systematically derive all the effective interactions between
$\mathbf{S}_{1}$ and $\mathbf{S}_{2}$ mediated by the conduction
electrons using second-order perturbation theory in $J_{K}$, and
without resorting to any specific model. In the process, not only
the RKKY exchange is obtained, but also Dzyaloshinskii-Moriya and
anisotropic tensor interactions.

We start from the full partition function of the system, which formally
writes as 
\begin{align}
Z & =\text{Tr}\left\{ e^{-\beta\left(H_{0}+\sum_{j}H_{K}\left(j\right)\right)}\right\} ,\nonumber \\
 & =\text{Tr}_{S}\ \text{Tr}_{\psi}\left\{ e^{-\beta\left(H_{0}+\sum_{j}H_{K}\left(j\right)\right)}\right\} ,\label{eq:Z_partial}
\end{align}
where $\beta=1/T$ (here we have assumed $k_{B}=1$). In Eq. (\ref{eq:Z_partial})
we have split the total trace into partial traces over fermionic (noted
as $\text{Tr}_{\psi}$) and spin (noted as $\text{Tr}_{S}$) degrees
of freedom. This allows to define the quantity $Z_{S}\equiv\text{Tr}_{\psi}\left\{ e^{-\beta\left(H_{0}+\sum_{j}H_{K}\left(j\right)\right)}\right\} $,
where the partial trace over the electrons is taken considering a
particular ``frozen'' configuration of the spins $\mathbf{S}_{1}$
and $\mathbf{S}_{2}$. In the zero-temperture limit $\beta\rightarrow\infty$,
this quantity allows to define an effective spin Hamiltonian where
the electronic degrees of freedom have been integrated out 
\begin{align}
Z_{S} & =e^{-\beta H_{\text{eff}}\left[\mathbf{S}_{1},\mathbf{S}_{2}\right]}\;\left(\text{for }\beta\rightarrow\infty\right).\label{eq:Z_S1S2_Heff}
\end{align}
Using the path-integral formalism \cite{negele}, $Z_{S}$ can be
expressed as 
\begin{align*}
Z_{S} & =\int\mathcal{D}\left[\bar{c},c\right]\ e^{-\mathcal{S}_{0}\left[\bar{c},c\right]-\sum_{j}\mathcal{S}_{K,j}\left[\mathbf{S}_{j},\bar{c},c\right]},
\end{align*}
where $\bar{c},c$ are Grassmann variables and $\mathcal{S}_{0}\left[\bar{c},c\right]$
and $\mathcal{S}_{K,j}\left[\mathbf{S}_{j},\bar{c},c\right]$ are,
respectively 
\begin{align}
\mathcal{S}_{0}\left[\bar{c},c\right] & =\sum_{\mathbf{k},n}\int_{0}^{\beta}d\tau\ \bar{c}_{\mathbf{k},n}\left(\tau\right)\left(\partial_{\tau}-\epsilon_{\mathbf{k},n}^{\left(0\right)}\right)c_{\mathbf{k},n}\left(\tau\right),\label{eq:S0}
\end{align}
the Euclidean action of the unperturbed Pb(110), expressed as an integral
over Matsubara time $\tau$ in the interval $\left[0,\beta\right]$,
and 
\begin{align}
\mathcal{S}_{K,j}\left[\mathbf{S}_{j},\bar{c},c\right] & =J_{K}\mathbf{S}_{j}\sum_{s,s^{\prime}}\int_{0}^{\beta}d\tau\Psi_{s}^{\dagger}\left(\mathbf{r}_{j},\tau\right)\frac{\hat{\boldsymbol{\sigma}}_{ss^{\prime}}}{2}\Psi_{s^{\prime}}\left(\mathbf{r}_{j},\tau\right),\label{eq:SKj}
\end{align}
is the Euclidean action of the \textit{s-d} interaction. The advantage
of the path-integral formalism is that it allows to express $Z_{S}$
as a series expansion in powers of $J_{K}$ as 
\begin{align}
Z_{S} & =Z_{0}\sum_{m=0}^{\infty}\frac{1}{m!}\left\langle \left(\sum_{j}\mathcal{S}_{K,j}\left[\mathbf{S}_{j},\bar{c},c\right]\right)^{m}\right\rangle _{0},\label{eq:Z_expansion}
\end{align}
where the notation $\left\langle A\right\rangle _{0}$ means the average
of operator $A$ with respect to the action $\mathcal{S}_{0}$, i.e.,
$\left\langle A\right\rangle _{0}=\int\mathcal{D}\left[\bar{c},c\right]\ e^{-\mathcal{S}_{0}\left[\bar{c},c\right]}A_{0}/Z_{0}$,
with $Z_{0}=\int\mathcal{D}\left[\bar{c},c\right]\ e^{-\mathcal{S}_{0}\left[\bar{c},c\right]}$
the partition function of unperturbed electrons in Pb(110).

The above Eqs. (\ref{eq:Z_partial})-(\ref{eq:Z_expansion}) are formally
exact, but in order to make progress we need to introduce a truncation
in the infinite series in Eq. (\ref{eq:Z_expansion}), assuming $J_{K}\rightarrow0$.
At second order, and introducing a subsequent cumultant expansion\cite{mahan},
the quantity $Z_{S}$ can be approximated as 
\begin{align}
Z_{S} & \approx e^{\frac{1}{2}\left\langle \left(\mathcal{S}_{K,1}\left[\mathbf{S}_{1},\bar{c},c\right]+\mathcal{S}_{K,2}\left[\mathbf{S}_{2},\bar{c},c\right]\right)^{2}\right\rangle _{0}},\label{eq:Z_cumulant}
\end{align}
(note that the first order term in (\ref{eq:Z_expansion}) has vanished
due to the time-reversal symmetry of the Pb(110) conduction band).
Comparing Eqs. (\ref{eq:Z_S1S2_Heff}) and (\ref{eq:Z_cumulant}),
we obtain the precise analytical form for the effective spin Hamiltonian
at second order in $J_{K}$: 
\begin{align}
H_{\text{eff}}\left[\mathbf{S}_{1},\mathbf{S}_{2}\right] & =-\frac{1}{2\beta}\lim_{\beta\rightarrow\infty}\left\langle \left(S_{K,1}\left[\mathbf{S}_{1},\bar{c},c\right]+S_{K,2}\left[\mathbf{S}_{2},\bar{c},c\right]\right)^{2}\right\rangle _{0},\label{eq:Heff_specific}
\end{align}
where the conduction-electrons of the Pb(110) band has been integrated
out. The effective Hamiltonian can be expressed as 
\begin{align}
H_{\text{eff}}\left[\mathbf{S}_{1},\mathbf{S}_{2}\right]= & \lim_{\beta\rightarrow\infty}\frac{J_{K}^{2}}{8}\frac{1}{\beta}\sum_{l}\sum_{i,j=1,2}\nonumber \\
 & \text{tr}\left\{ \left(\mathbf{S}_{i}.\hat{\boldsymbol{\sigma}}\right)\hat{\mathbf{g}}_{0}\left(\mathbf{r}_{i},\mathbf{r}_{j},i\nu_{l}\right)\left(\mathbf{S}_{j}.\hat{\boldsymbol{\sigma}}\right)\hat{\mathbf{g}}_{0}\left(\mathbf{r}_{j},\mathbf{r}_{i},i\nu_{l}\right)\right\} ,\label{eq:Heff_compact}
\end{align}
where tr\{...\} is the usual trace of a matrix, and $\hat{\mathbf{g}}_{0}\left(\mathbf{r}_{i},\mathbf{r}_{j},i\nu_{l}\right)$
is the matrix of the unperturbed conduction-electron propagators in
Pb(110) 
\begin{align}
\hat{\mathbf{g}}_{0}\left(\mathbf{r}_{i},\mathbf{r}_{j},i\nu_{l}\right) & =\left(\begin{array}{cc}
g_{0}^{\left(\uparrow\uparrow\right)}\left(\mathbf{r}_{j},\mathbf{r}_{i},i\nu_{l}\right) & g_{0}^{\left(\uparrow\downarrow\right)}\left(\mathbf{r}_{j},\mathbf{r}_{i},i\nu_{l}\right)\\
g_{0}^{\left(\downarrow\uparrow\right)}\left(\mathbf{r}_{j},\mathbf{r}_{i},i\nu_{l}\right) & g_{0}^{\left(\downarrow\downarrow\right)}\left(\mathbf{r}_{j},\mathbf{r}_{i},i\nu_{l}\right)
\end{array}\right),\label{eq:g0_matrix}
\end{align}
with matrix elements 
\begin{align}
g_{0}^{\left(ss^{\prime}\right)}\left(\mathbf{r}_{j},\mathbf{r}_{i},i\nu_{l}\right) & =\sum_{\mathbf{k},n}\frac{\psi_{\mathbf{k}n}^{(s)}\left(\mathbf{r}_{j}\right)\psi_{\mathbf{k}n}^{*(s^{\prime})}\left(\mathbf{r}_{i}\right)}{i\nu_{l}-\epsilon_{\mathbf{k},n}^{\left(0\right)}}\;\left(s,s^{\prime}=\left\{ \uparrow,\downarrow\right\} \right).\label{eq:g0_ssprime}
\end{align}
In this expression we have introduced the fermionic Matsubara frequencies
$i\nu_{l}=2\pi i\left(l+\frac{1}{2}\right)/\beta$. Physically, the
Green's function $g_{0}^{\left(ss^{\prime}\right)}\left(\mathbf{r}_{j},\mathbf{r}_{i},i\nu_{l}\right)$
measures the probability that an electron created at $\mathbf{r}_{i}$
with spin $s^{\prime}$ arrives at $\mathbf{r}_{j}$ with spin $s$
in the unperturbed surface of Pb(110). Note that in absence of SOC,
the spin-projection labels $s$ and $s^{\prime}$ would be good quantum
numbers and therefore the off-diagonal elements would vanish. Moreover,
due to the SU(2) symmetry in the absence of SOC and externally applied
magnetic fields, $g_{0}^{\left(\uparrow\uparrow\right)}\left(\mathbf{r}_{j},\mathbf{r}_{i},i\nu_{l}\right)=g_{0}^{\left(\downarrow\downarrow\right)}\left(\mathbf{r}_{j},\mathbf{r}_{i},i\nu_{l}\right)$
and therefore the matrix $\hat{\mathbf{g}}_{0}\left(\mathbf{r}_{i},\mathbf{r}_{j},i\nu_{l}\right)$
would be a scalar proportional to the unit matrix. In what follows,
we introduce a more convenient representation of the propagator matrix
(\ref{eq:g0_matrix}) in terms of the $2\times2$ Pauli matrices \cite{Imamura2004}
\begin{align}
\hat{\mathbf{g}}_{0}\left(\mathbf{r}_{i},\mathbf{r}_{j},i\nu_{l}\right) & =g_{0}^{0}\left(\mathbf{r}_{i},\mathbf{r}_{j},i\nu_{l}\right)\mathbf{1}_{2\times2}+g_{0}^{x}\left(\mathbf{r}_{i},\mathbf{r}_{j},i\nu_{l}\right)\hat{\sigma}_{x}\nonumber \\
 & +g_{0}^{y}\left(\mathbf{r}_{i},\mathbf{r}_{j},i\nu_{l}\right)\hat{\sigma}_{y}+g_{0}^{z}\left(\mathbf{r}_{i},\mathbf{r}_{j},i\nu_{l}\right)\hat{\sigma}_{z},\label{eq:g0_matrix_Pauli}
\end{align}
where the new propagators $g_{0}^{k}\left(\mathbf{r}_{i},\mathbf{r}_{j},i\nu_{l}\right)$
(with $k=\left\{ 0,x,y,z\right\} $) are linear combinations of the
propagators (\ref{eq:g0_ssprime}), which allow to readily evaluate
the trace in Eq.(\ref{eq:Heff_compact}), and express the Hamiltonian
as 
\begin{align}
H_{\text{eff}}\left[\mathbf{S}_{1},\mathbf{S}_{2}\right] & =J_{\text{RKKY}}\left(\mathbf{r}_{1},\mathbf{r}_{2}\right)\mathbf{S}_{1}.\mathbf{S}_{2}+\mathbf{D}_{\text{DM}}\left(\mathbf{r}_{1},\mathbf{r}_{2}\right).\left(\mathbf{S}_{1}\times\mathbf{S}_{2}\right)\nonumber \\
 & +2\mathbf{S}_{1}.\mathbf{T}\left(\mathbf{r}_{1},\mathbf{r}_{2}\right).\mathbf{S}_{2}+\mathbf{S}_{1}.\mathbf{T}\left(\mathbf{r}_{1},\mathbf{r}_{1}\right).\mathbf{S}_{1}\nonumber \\
 & +\mathbf{S}_{2}.\mathbf{T}\left(\mathbf{r}_{2},\mathbf{r}_{2}\right).\mathbf{S}_{2}.\label{eq:Heff_compact2}
\end{align}
Here we have defined the scalar RKKY exchange-interaction as 
\begin{align}
J_{\text{RKKY}}\left(\mathbf{r}_{1},\mathbf{r}_{2}\right) & =\frac{J_{K}^{2}}{2}\frac{1}{\beta}\sum_{l}\biggl[g_{0}^{0}\left(\mathbf{r}_{1},\mathbf{r}_{2},i\nu_{l}\right)g_{0}^{0}\left(\mathbf{r}_{2},\mathbf{r}_{1},i\nu_{l}\right)\nonumber \\
 & -\sum_{j=\left\{ x,y,z\right\} }g_{0}^{j}\left(\mathbf{r}_{1},\mathbf{r}_{2},i\nu_{l}\right)g_{0}^{j}\left(\mathbf{r}_{2},\mathbf{r}_{1},i\nu_{l}\right)\biggr],\label{eq:J_RKKY}
\end{align}
The next term in Eq. (\ref{eq:Heff_compact2}) corresponds to the
Dzyaloshinskii-Moriya interaction 
\begin{align}
D_{\text{DM}}^{j}\left(\mathbf{r}_{1},\mathbf{r}_{2}\right) & =i\frac{J_{K}^{2}}{2}\frac{1}{\beta}\sum_{l}\left[g_{0}^{0}\left(\mathbf{r}_{1},\mathbf{r}_{2},i\nu_{l}\right)g_{0}^{j}\left(\mathbf{r}_{2},\mathbf{r}_{1},i\nu_{l}\right)\right.\nonumber \\
 & \left.-g_{0}^{j}\left(\mathbf{r}_{1},\mathbf{r}_{2},i\nu_{l}\right)g_{0}^{0}\left(\mathbf{r}_{2},\mathbf{r}_{1},i\nu_{l}\right)\right]\;\left(j=\left\{ x,y,z\right\} \right),\label{eq:D_DM}
\end{align}
which is an anisotropic vector interaction. Finally, the last terms
are anisotropic tensor interactions of the form 
\begin{align}
T^{jk}\left(\mathbf{r}_{1},\mathbf{r}_{2}\right) & =\frac{J_{K}^{2}}{4}\frac{1}{\beta}\sum_{l}\left[g_{0}^{j}\left(\mathbf{r}_{1},\mathbf{r}_{2},i\nu_{l}\right)g_{0}^{k}\left(\mathbf{r}_{2},\mathbf{r}_{1},i\nu_{l}\right)\right.\nonumber \\
 & \left.+g_{0}^{j}\left(\mathbf{r}_{2},\mathbf{r}_{1},i\nu_{l}\right)g_{0}^{k}\left(\mathbf{r}_{1},\mathbf{r}_{2},i\nu_{l}\right)\right],\;\left(j,k=\left\{ x,y,z\right\} \right)\label{eq:T_iso}
\end{align}
which generalize the Ising and the single-ion magneto-crystalline
contributions. Note that in (\ref{eq:Heff_compact2}) we have neglected
the RKKY self-interaction terms $J_{\text{RKKY}}\left(\mathbf{r}_{j},\mathbf{r}_{j}\right)\mathbf{S}_{j}^{2}$,
since they are only a renormalization of the energy. %

Although the three contributions Eqs. (\ref{eq:J_RKKY})-(\ref{eq:T_iso})
are of the same order $\mathcal{O}\left(J_{K}^{2}\right)$, their
relative magnitude strongly depends on the magnitude of the Rashba
SOC parameter $\alpha_{R}$. This can be understood directly at the
level of the propagators $g_{0}^{\left(x,y\right)}\left(\mathbf{r}_{1},\mathbf{r}_{2},i\nu_{l}\right)$
appearing in these expressions, which are directly proportional to
the SU(2) symmetry-breaking terms in the Hamiltonian, as shown in
previous works\cite{Imamura2004,Bouaziz2017}. Then, it is easy to
see that 
\begin{align*}
J_{\text{RKKY}}\left(\mathbf{r}_{1},\mathbf{r}_{2}\right) & \sim\mathcal{O}\left(1\right),\\
\left|\mathbf{D}_{\text{DM}}\left(\mathbf{r}_{1},\mathbf{r}_{2}\right)\right| & \sim\mathcal{O}\left(\alpha_{R}\right),\\
\left\Vert \mathbf{T}\left(\mathbf{r}_{1},\mathbf{r}_{2}\right)\right\Vert  & \sim\mathcal{O}\left(\alpha_{R}^{2}\right),
\end{align*}

\section{Methods and technical considerations}
\label{sec:methods}

A technical point in the derivation of the RKKY interaction Eq. (\ref{eq:J_RKKY})
concerns the sum over the band index $n$, whose convergence is very
slow. In principle, this sum runs over an infinite number of bands,
but in practice must be limited by a cutoff energy $E_{c}$ that ensures
the convergence of the involved quantities. Due to its poor convergence
properties, for a reasonable accuracy in the value of $J_{\text{RKKY}}\left(\mathbf{r}_{1},\mathbf{r}_{2}\right)$
an unfeasible large value of $E_{c}$ would be necessary (even for
values of the order of $E_{c}=$100 eV, errors would still be over
100\%). However, it is physically expected that above a certain $E_{c}$
the system wavefunctions are indistinguishable from the corresponding
free-electron wavefunctions at the same energy. In other words, at
sufficiently high energies $\epsilon_{\mathbf{k},n}^{\left(0\right)}$,
it is expected that 
\begin{align}
\epsilon_{\mathbf{k},n}^{\left(0\right)} & \approx\epsilon_{\mathbf{k}}^{\text{free}}=\frac{\hbar^{2}\left(\mathbf{k}+\mathbf{G}_{n}\right)^{2}}{2m},\\
\psi_{\mathbf{k}n}^{(s)}\left(\mathbf{r}\right) & \approx\frac{e^{i\left(\mathbf{k}+\mathbf{G}_{n}\right).\mathbf{r}}}{\sqrt{V}},
\end{align}
with $\mathbf{G}_{n}$ a reciprocal-lattice vector. Taking this into
account, the RKKY exchange interaction Eq. (\ref{eq:J_RKKY}) at $T=0$
can then be recast as 
\begin{align}
J_{\text{RKKY}}\left(\mathbf{r}_{1},\mathbf{r}_{2}\right) & =J_{\text{RKKY}}^{0}\left(\mathbf{r}_{1},\mathbf{r}_{2}\right)+J_{\text{RKKY}}^{\text{free}}\left(\mathbf{r}_{1},\mathbf{r}_{2}\right),\label{eq:rkkycorr1}
\end{align}
where 
\begin{widetext}
\begin{align}
J_{\text{RKKY}}^{0}\left(\mathbf{r}_{1},\mathbf{r}_{2}\right) & =\frac{J_{K}^{2}}{4}\sum_{s=\left\{ \uparrow,\downarrow\right\} }\sum_{\mathbf{k},n}^{\epsilon_{\mathbf{k}n}<E_{F}}\times\frac{1}{V}\sum_{\mathbf{k^{\prime}},n^{\prime}}^{E_{F}<\epsilon_{\mathbf{k^{\prime}}n^{\prime}}<E_{c}}\frac{1}{\epsilon_{\mathbf{k}n}^{\left(0\right)}-\epsilon_{\mathbf{k^{\prime}}n^{\prime}}^{\left(0\right)}}\nonumber \\
 & \times\left[\psi_{\mathbf{k}n}^{(s)}\left(\mathbf{r}_{1}\right)\psi_{\mathbf{k}n}^{*(s)}\left(\mathbf{r}_{2}\right)\psi_{\mathbf{k^{\prime}}n^{\prime}}^{\left(\bar{s}\right)}\left(\mathbf{r}_{2}\right)\psi_{\mathbf{k^{\prime}}n^{\prime}}^{*\left(\bar{s}\right)}\left(\mathbf{r}_{1}\right)-\psi_{\mathbf{k}n}^{(s)}\left(\mathbf{r}_{1}\right)\psi_{\mathbf{k}n}^{*(\bar{s})}\left(\mathbf{r}_{2}\right)\psi_{\mathbf{k^{\prime}}n^{\prime}}^{\left(\bar{s}\right)}\left(\mathbf{r}_{2}\right)\psi_{\mathbf{k^{\prime}}n^{\prime}}^{*\left(s\right)}\left(\mathbf{r}_{1}\right)\right],\\
J_{\text{RKKY}}^{\text{free}}\left(\mathbf{r}_{1},\mathbf{r}_{2}\right) & =\frac{J_{K}^{2}}{4}\sum_{s=\left\{ \uparrow,\downarrow\right\} }\sum_{\mathbf{k},n}^{\epsilon_{\mathbf{k}n}<E_{F}}\frac{1}{V}\sum_{\mathbf{k^{\prime}}}^{E_{c}<\epsilon_{\mathbf{k^{\prime}}}}\left[\psi_{\mathbf{k}n}^{(s)}\left(\mathbf{r}_{1}\right)\psi_{\mathbf{k}n}^{*(s)}\left(\mathbf{r}_{2}\right)-\psi_{\mathbf{k}n}^{(s)}\left(\mathbf{r}_{1}\right)\psi_{\mathbf{k}n}^{*(\bar{s})}\left(\mathbf{r}_{2}\right)\right]\times\frac{e^{i\mathbf{k}^{\prime}.\left(\mathbf{r}_{2}-\mathbf{r}_{1}\right)}}{\epsilon_{\mathbf{k}n}^{\left(0\right)}-\epsilon_{\mathbf{k}^{\prime}}^{\text{free}}},
\end{align}
\end{widetext}
where in this last equation we have dropped the index $n^{\prime}$
and let $\mathbf{k}^{\prime}$ run over the extended Brillouin zone.
$J_{\text{RKKY}}^{\text{free}}\left(\mathbf{r}_{1},\mathbf{r}_{2}\right)$
is then a free-electron correction term that can be analytically integrated
and that only depends on states below the Fermi energy and on the
numerical value of $E_{c}$. In this way, instead of summing over
a large number of bands, Eq. (\ref{eq:rkkycorr1}) only needs to be
evaluated until convergence with respect to $E_{c}$ is attained.

The wavefunctions $\psi_{\mathbf{k},n}^{\left(s\right)}\left(\mathbf{r}_{j}\right)$
required to calculate the interactions Eqs. (\ref{eq:J_RKKY})-(\ref{eq:T_iso})
for both impurities were obtained from DFT calculations performed
by using the VASP code\cite{vaspcode1,vaspcode2,vaspcode3}. The
Pb (110) substrate was modeled as a periodic slab consisting of an
\textcolor{black}{1$\times$1 surface} with $N$ layers of Pb atoms
and a vacuum layer of 15 Å\ to avoid coupling between surfaces in
different periodic cells. Three top layers were allowed to relax while
the other ones where kept fixed at their bulk lattice coordinates.
Ionic forces were converged to be lower than 0.01 eV/{Å}, with a
cutoff of 150 eV and using a Monhorst-Pack\cite{monkhorstpack} grid
of 10$\times$10$\times$1 $k$-points. All calculations were performed
within the PAW method\cite{vasppaw} and using PBEsol exchange-correlation
functional\cite{pbesol}. Since the interactions (\ref{eq:J_RKKY})-(\ref{eq:T_iso})
are highly dependent on the accuracy of the wavefunctions $\psi_{\mathbf{k},n}^{\left(s\right)}\left(\mathbf{r}_{j}\right)$,
a more stringent convergence condition and a larger number of empty
bands were needed in their calculation. In order to obtain errors
within 5\% we used 120 empty bands and 1600 $k$-points. Calculations
were performed with and without atomic spin-orbit interaction.

\section{Results}

\label{sec:results}

\subsection{Band Structure of Pb(110) and estimation of Rashba coupling parameter}

\label{subsec:band_structure_pb110}

\begin{figure}
\includegraphics[scale=0.6]{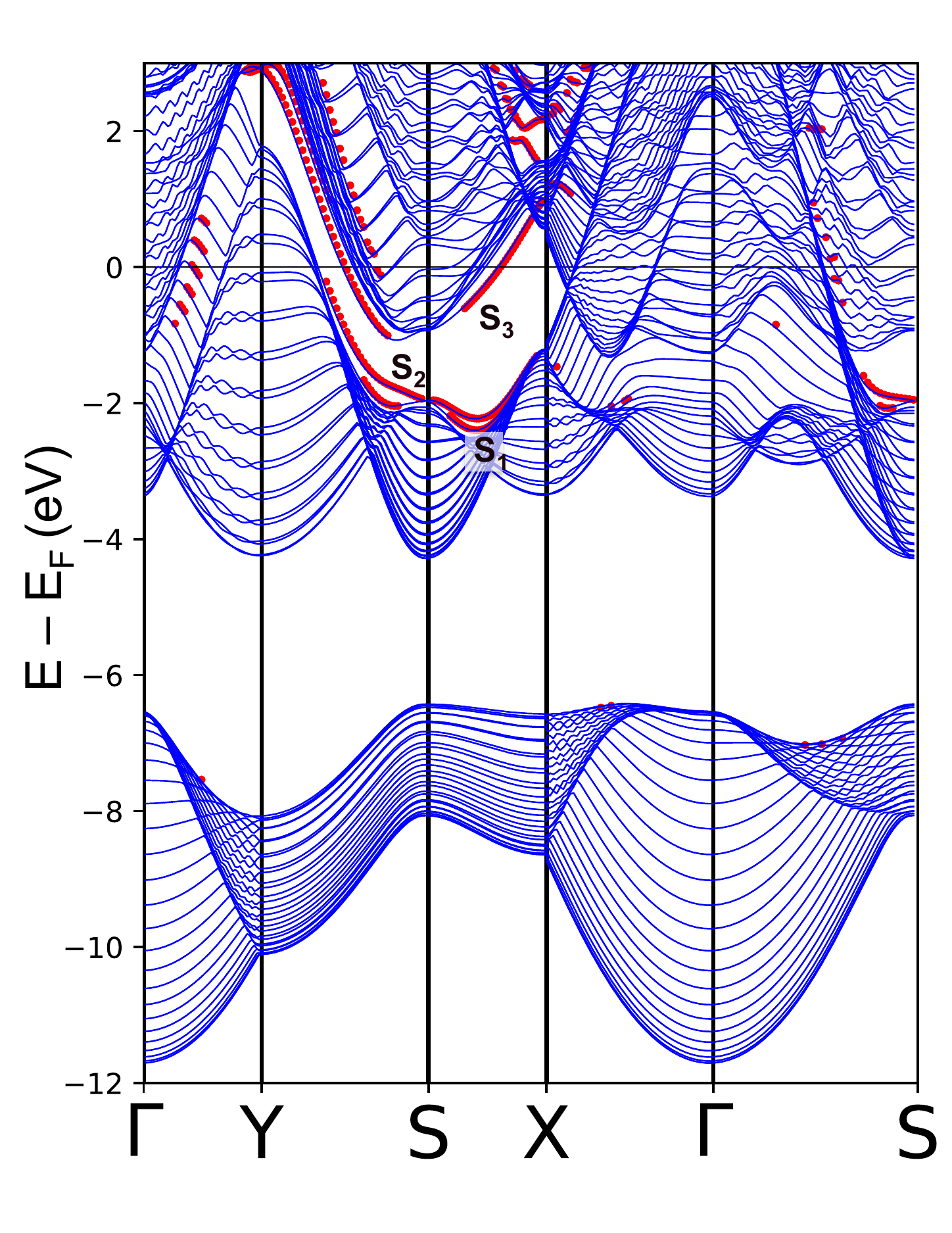}\caption{(Color online) Calculated band structure for a 29-layer Pb(110) slab
along high symmetry paths of the Brilluoin zone. Surface and resonant
states, represented by red dots, have been obtained by requiring the
density for a given state projected onto the surface layers (top and
bottom) be greater than 0.3. The surface states are labeled as S$_{1}$, S$_{2}$ and S$_{3}$.}
\label{fig:pb110_bands} 
\end{figure}

The band structure of Pb(110) bulk and surface was first studied by
Würde \textit{et al}\cite{wurde94} using the empirical tight-binding
method (ETBM) combined with the scattering-theory method to determine
the different surface and resonant states. Given the large atomic
number of Pb, the effect of the spin-orbit interaction cannot be neglected
for this system, and needs to be taken into account. In the present
work the Pb(110) band structure was obtained by self-consistent DFT
calculations by including spin-orbit coupling and by also considering
relaxation effects on the (110) surface. In Figure \ref{fig:pb110_bands}
we show the calculated band structure for a Pb(110) slab with $N=29$
layers. The surface and resonant states (red dots) are identifed as
those for which the sum of the square projections onto the top and
bottom layers of the slab is greater than 30\%. These results are
in excellent agreement with the ETBM calculations of Würde \textit{et
al}. The large gap in the range between -6.5 to -4 eV corresponds
to the marked energy difference between $s$ and $p$ levels in bulk
Pb. Surface states (denoted as S$_{1}$, S$_{2}$ and S$_{3}$ in
Fig. \ref{fig:pb110_bands}) are mainly localized near the edges of
the gaps arising along the X and Y directions, and extend between
-2.4 and 3 eV. Bands S$_{1}$ and S$_{2}$ (S$_{3}$) consist mostly
of $p$-states parallel (perpendicular) to the 110 surface. While
the gap opening at point S for $E$=-2 eV is a consequence of spin-orbit
coupling, the splitting of bands S$_{1}$ and S$_{2}$ arises from
the Rashba shift at S, generated by the breaking of the symmetry along
$z$. Indeed, by using the $\mathbf{k}\cdot\mathbf{p}$ approximation
to fit bands S$_{1}$ and S$_{2}$, it is possible to estimate the
Rashba parameter to be $\alpha_{R}=0.97$eV.{Å}.

The excellent agreement with the results by Würde \textit{et al} justifies
the use of the Kohn-Sham orbitals as the wavefunctions $\psi_{\mathbf{k}n}^{(s)}\left(\mathbf{r}_{j}\right)$
appearing in the expression of the unperturbed propagator Eq. (\ref{eq:g0_ssprime})
and in the calculation of the magnetic interactions Eqs. (\ref{eq:J_RKKY})-(\ref{eq:T_iso}).

\subsection{Convergence of the RKKY interaction}

\label{subsec:convergence}

\begin{figure}
\includegraphics[scale=0.5]{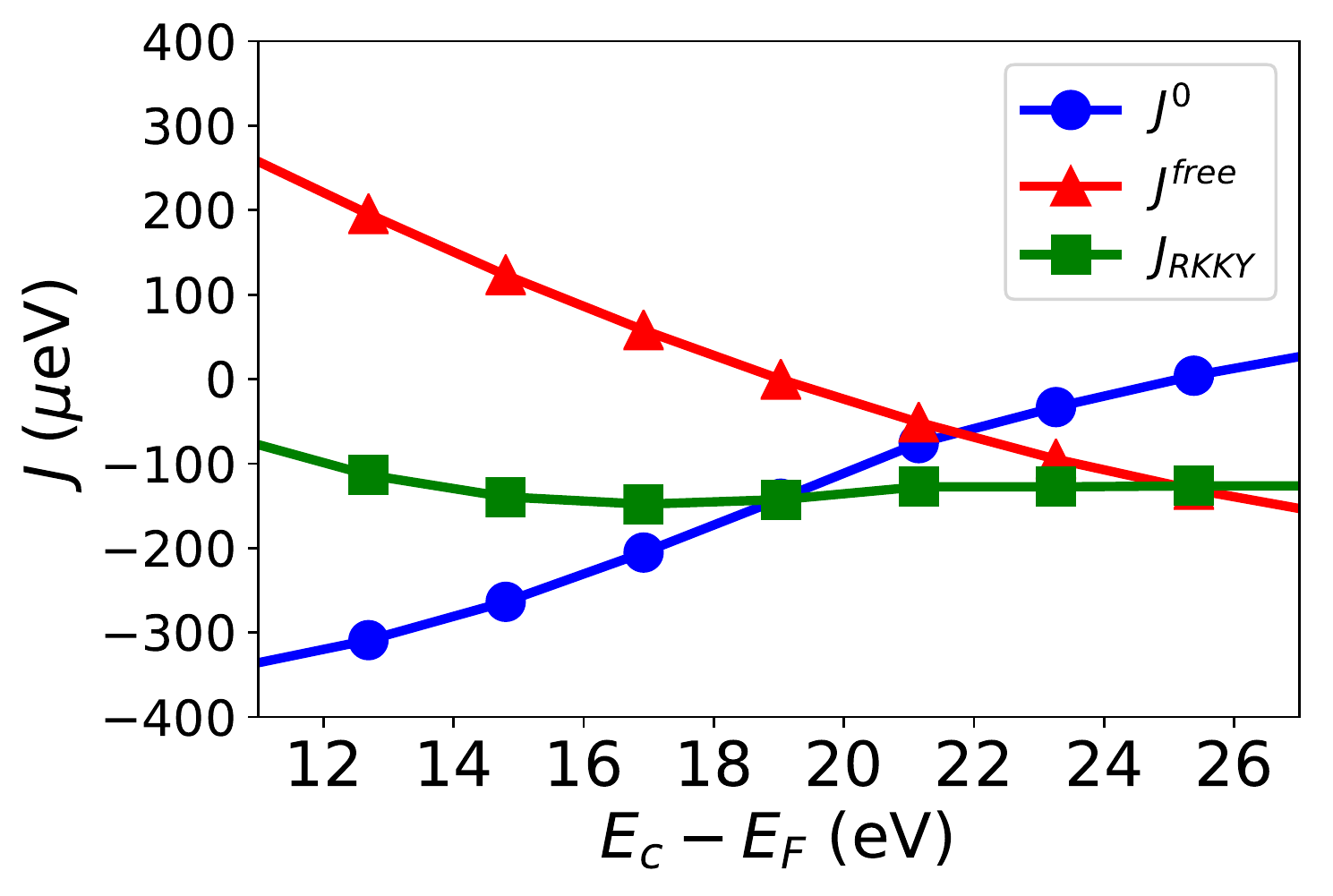}\caption{(Color online) Convergence of the RKKY interaction $J_{\text{RKKY}}$
(green squares) computed as in Eq. (\ref{eq:rkkycorr1}), as a function
of the plane-wave cutoff energy $E_{c}-E_{F}$. Convergence is attained
when the free-electron correction $J^{\text{free}}$
(red triangles) exactly mirrors the uncorrected RKKY term $J^{0}$
(blue circles), for $E_{c}=$22 eV above $E_{F}$.}
\label{fig2}
\end{figure}

In order to ensure the convergence of the RKKY interaction with respect
to the cutoff $E_{c}$, we evaluated the function $J_{\text{RKKY}}\left(\mathbf{r}_{1},\mathbf{r}_{2}\right)$
as in Eq. (\ref{eq:rkkycorr1}). Since the magnitude of the correction
term $J_{\text{RKKY}}^{\text{free}}\left(\mathbf{r}_{1},\mathbf{r}_{2}\right)$
varies inversely with the distance between the two impurities, impurities
located at larger distances would require smaller values of $E_{c}$
in order to converge. For this reason the cutoff energy needs to be
optimized for the minimum distance considered, which in our case corresponds
to one half of the $b$ lattice parameter (or, in other words, when
$\left|\mathbf{r_{2}-r_{1}}\right|=0.5 b$). In Fig. \ref{fig2} we
display the total coupling $J_{\text{RKKY}}\equiv J_{\text{RKKY}}\left(0, 0.5 b\ \hat{\mathbf{y}}\right)$
together with $J^{0}\equiv J_{\text{RKKY}}^{0}\left(0,0.5 b\ \hat{\mathbf{y}}\right)$
and its correction $J^{\text{free}}\equiv J_{\text{RKKY}}^{\text{free}}\left(0,0.5 b\ \hat{\mathbf{y}}\right)$,
corresponding to a Pb(110) slab with $N=11$ Pb layers for different
values of the cutoff energy $E_{c}$, taken with respect to $E_{F}$.
Two observations become apparent in the plot. On the one hand, the
slow decreasing rate of the correction terms (actually, oscillatory
and analogous to the integral sin/cosine functions), which makes them
impossible to neglect even for a very high energy cutoff. On the other
hand, we observe that the correction closely compensates the variation
of $J^{0}$ with respect to the cutoff. Despite the fact that for
small values of $E_{c}$ the free electron approximation is still
too crude and the errors large, for values of $E_{c}$ larger than
$E_{F}+20$ eV, the rate of variation of $J^{\text{free}}$ clearly
mirrors $J^{0}$. In this last case, the total $J_{\text{RKKY}}$
becomes flat and converges within a 2\% of error. The same calculations
were repeated for a $N=$17-layer slab, obtaining a similar cutoff
and the same values for $J_{\text{RKKY}}$ (always within 2\% error)
as for the 11-layer case. In the light of these results and for the
sake of simplicity, the rest of our calculations were performed for
an 11-layer slab using a value $E_{c}=$22 eV.

\subsection{Magnetic Interactions}

\label{subsec:magnetic_interactions}

As mentioned earlier, the free parameter $J_{K}$ must be determined
in order to compute the magnetic interactions in Eqs. (\ref{eq:J_RKKY})-(\ref{eq:T_iso}).
Here we estimated $J_{K}$ by combining our DFT calculations with
results from STM experiments studying the Kondo resonance in self-assembled
metal-organic complexes (i.e., MnPc molecules) deposited on Pb(111)\cite{Franke_2011}.
In these experiments the weak character of the magnetic interactions
between the absorbed molecules is established by the close proximity
of the measured subgap quasiparticle peaks (i.e., the ``Shiba peaks'') to the superconducting gap. Furthermore,
in these works the Kondo temperature $T_{K}$ obtained from the Fano
resonance near $E_{F}$ ranges from 200 K to 400 K. If we consider
an intermediate value $T_{K}\approx$ 300 K, $J_{K}$ can then be
extracted from the Kondo temperature formula $k_{B}T_{K}=We^{-1/\rho_\text{3D}J_{K}}$\cite{hewson}.
In this equation the value of the local density of states per unit
volume at the Fermil level, $\rho_\text{3D}$, is obtained from DFT and
corresponds to Pb's located at the (110) surface. Within this model
(which assumes a simplified rectangular flat band) the half bandwidth of the
conduction states is calculated as $W=1/(2\rho_\text{3D}V_{atom}),$ with
$V_{atom}$ being the atomic volume for Pb. From this we obtain $\rho_\text{3D}=0.01\,\mathrm{eV^{-1}}\mathrm{\mathring{A}^{-3}}$
and $J_{K}=10.80\,\mathrm{eV.\mathring{A}^{3}},$thus yielding $\rho_\text{3D}J_{K}\approx0.1$,
which is consistent with our aforementioned assumption of weakly coupled
adsorbates (such as MnPc on a Pb surface).

\begin{figure}
\begin{centering}
\includegraphics[scale=0.43]{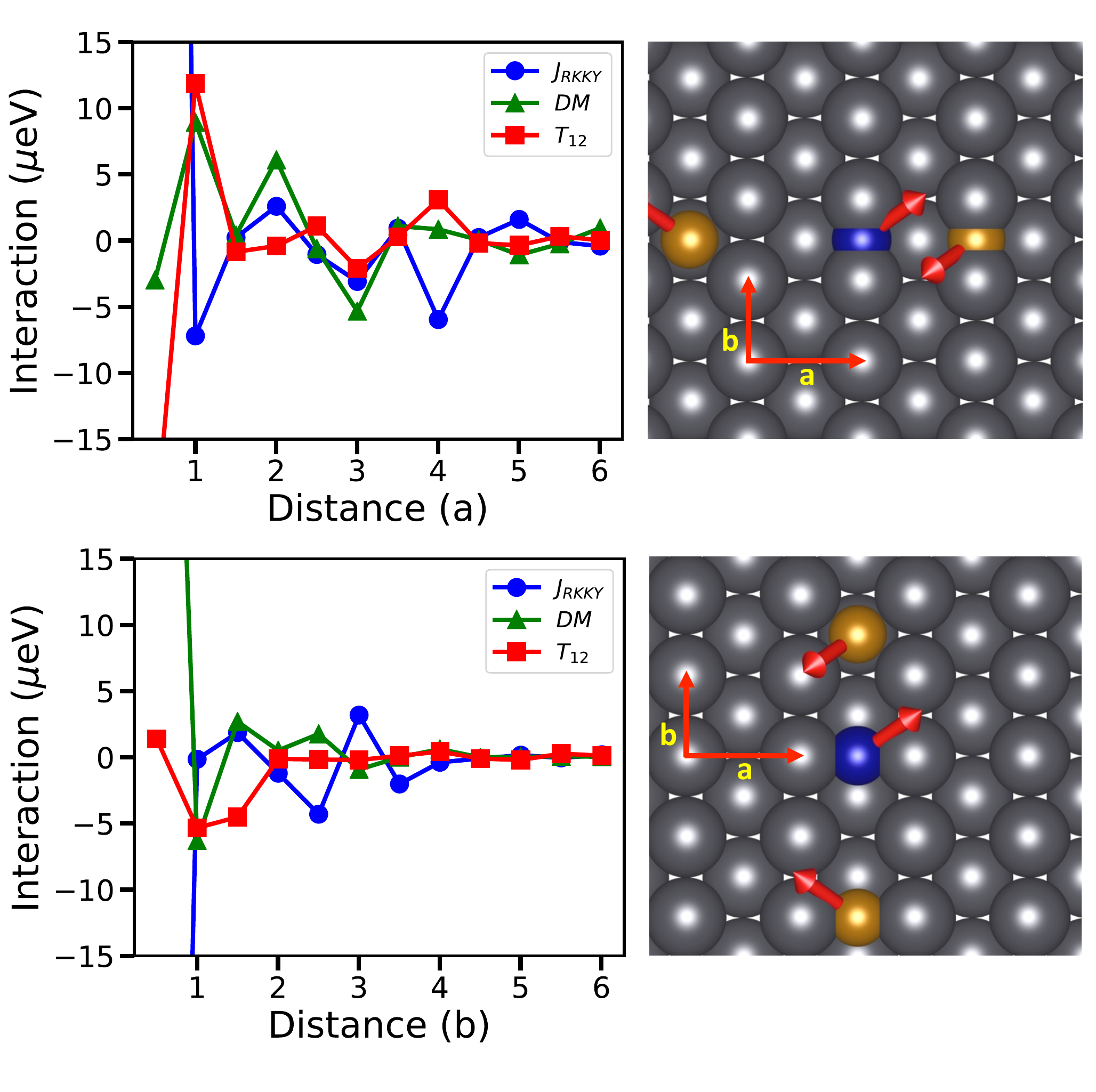}
\par\end{centering}
\caption{(Color online) Top (Bottom): RKKY, DM and $T_{12}$ magnetic interaction parameters for configurations
with the reference impurity (blue) placed at the middle of the shorter (longer) bridge between 
two Pb atoms on the (110) surface and the second impurity (yellow) at $n.a$/2 ($n.b$/2) lattice parameters
away along $a$($b$)-direction.}
\label{interactions1}
\end{figure}

\begin{figure}
\begin{centering}
\includegraphics[scale=0.43]{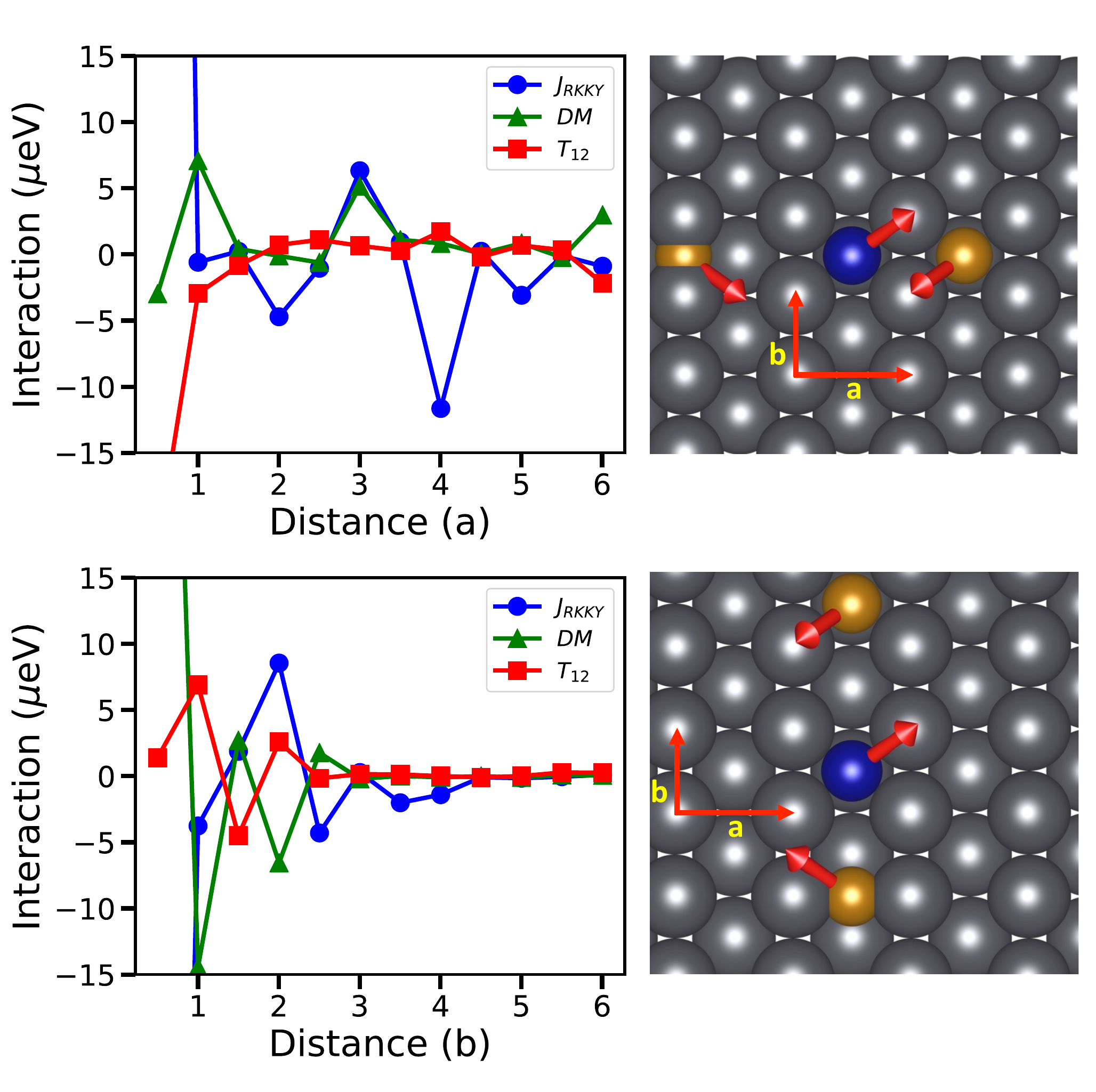}
\par\end{centering}
\caption{(Color online) Top (Bottom): RKKY, DM and $T_{12}$ magnetic interaction parameters for configurations
with the reference impurity (blue) placed at $(0.5a,0.5b)$  on the (110) surface and the second 
impurity (yellow) at $n.a$/2 ($n.b$/2) lattice parameters
away along $a$($b$)-direction.
}
\label{interactions2}
\end{figure}

Once the parameter $J_{K}$ has been obtained, the different magnetic
interactions in Eqs. (\ref{eq:J_RKKY})-(\ref{eq:T_iso}) were calculated
as a function of distance for impurities located along the $a$ and
$b$ directions on the Pb(110) surface. We considered two inequivalent
positions for each impurity at the surface: either locating it halfway on 
the bridge between two Pb atoms of the top layer or above a Pb atom 
belonging to the layer immediately below [$i.e.$ at $(0.5a,0.5b)$]. 
In Fig. \ref{interactions1} we plot the RKKY,
DM and anisotropic-tensor ($T_{12}$) interactions locating a reference impurity at
the bridge between two Pb's and letting the second impurity be located at either
inequivalent position, in such a way that the distance between impurities
is either an integer or half integer of one of the in-plane lattice parameters. Top and bottom
panels in Fig. \ref{interactions1} display the calculated interactions
for the reference impurity at a bridge location and the second impurity at different distances
along the $a$ and $b$ directions, respectively. In Fig. \ref{interactions2}
we repeated the calculations but this time locating the reference magnetic impurity 
at $(0.5a,0.5b)$. In spite of the fact that all interactions follow the expected
oscillatory behavior, their character departs significantly from the
smooth and monotonic decay of the classical RKKY. It is also worth
noticing the anisotropic character of the interactions: while we still
encounter large peaks for impurities located 4 lattice parameters apart
along $a$, interactions are significantly reduced for the same relative
distance along $b$. This behavior, which at first glance may seem
counterintuitive (since the impurities are at a closer distance along
$b$), can be qualitatively understood by noticing that the flatter
bands along $a$ give rise to a larger density of states along this
direction than along $b$, thus enhancing the magnitude of the interactions.

\begin{figure}
\includegraphics[scale=0.35]{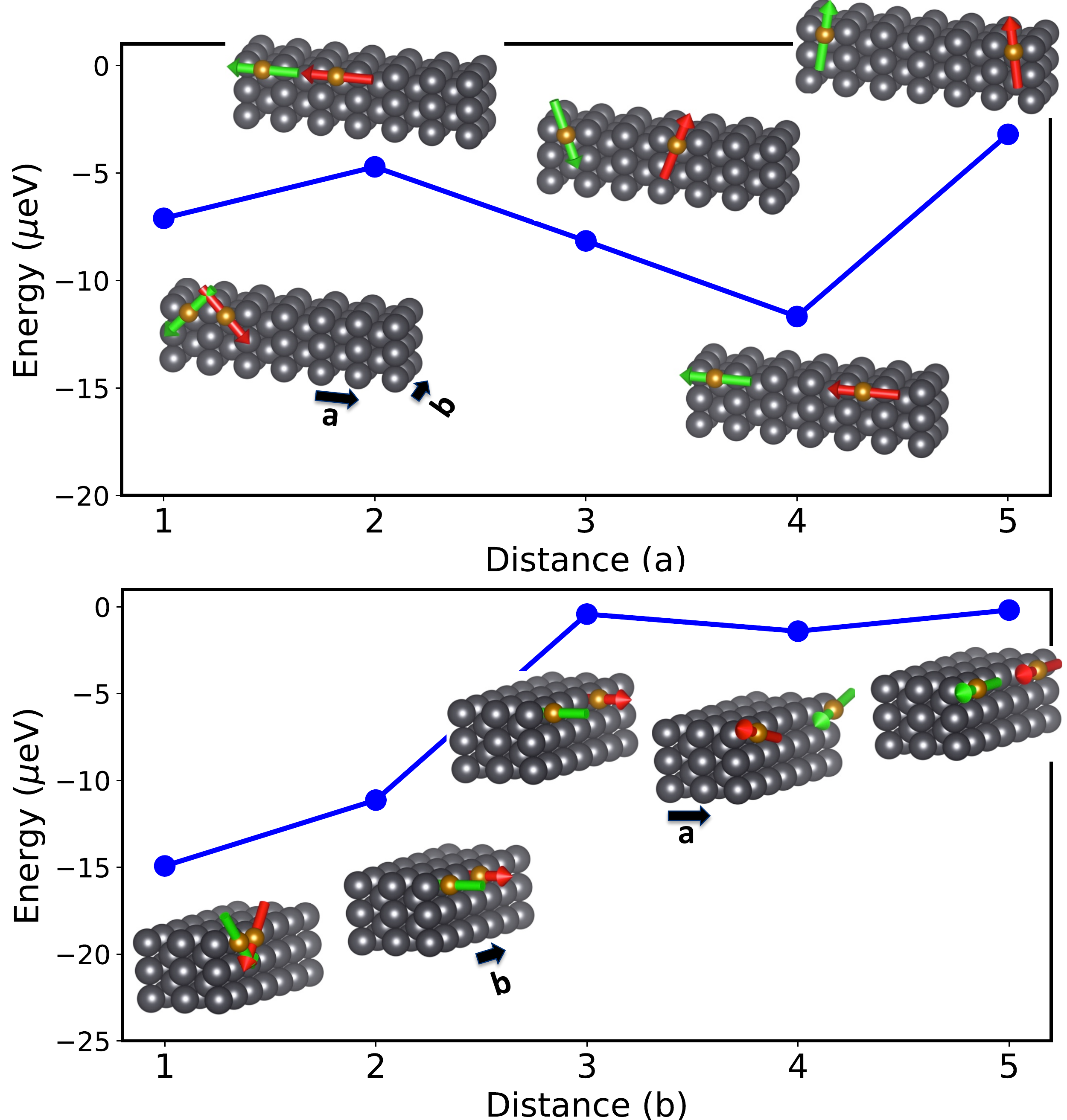}\caption{(Color online) Top panel: Ground state energies and their corresponding spin configurations
for the magnetic Hamiltonian of Equation \ref{eq:Heff_compact} with
a reference impurity (blue) located at $(0.5a,0.5b)$ interacting
with impurites (yellow) located at $(0.5na,0.5b)$ on the Pb (110)
surface. Bottom panel: Now the impurities interact along the $\mathbf{b}$-direction,
i.e., the second impurity is placed at $(0.5a,0.5nb)$ .}

\label{gsmagconf}
\end{figure}

With the calculated magnetic interactions ($J_\text{RKKY}$, DM, $T_{12}$)
it is then possible to search the spin configurations that minimize
the effective Hamiltonian Eq. \ref{eq:Heff_compact2}, i.e., we can
obtain the classical ground state configuration. In Fig. \ref{gsmagconf}\textcolor{black}{,
top and bottom panels,}\textcolor{red}{{} }we display these configurations
together with their corresponding energies for interacting spins located
at $\mathbf{r}_{1}=\left(0.5a,0.5b,0\right)$ and $\mathbf{r}_{2}=\left(\left(0.5+n\right)a,0.5b,0\right)$,
and at $\mathbf{r}_{1}=\left(0.5a,0.5b,0\right)$ and $\mathbf{r}_{1}=\left(0.5a,(0.5+nb),0\right)$,
respectively. Taking for instance impurities one lattice parameter
apart along the $a$ direction, we see from Fig. \ref{interactions2}
that in this case the dominant interaction is DM, thus favoring a
canted spin configuration as the one shown in Fig. \ref{gsmagconf}.
Analogously, when the impurites are a located two lattice parameters
apart along the $a$ direction, both DM and $T_{12}$ interactions nearly
vanish and the dominant interaction is RKKY, resulting in a collinear
spin configuration, which since $J_{\text{RKKY}}<0$ is ferromagnetic.
Along the direction $b$, the behavior of magnetic ground state energy
is approximately monotonic, and %the Despite the fact that the energy of the magnetic ground state is the lowest for impurities located one lattice parameter apart and along b, 
its value is abruptly reduced after $n=3b$ and beyond. Contrarily,
along the $a$ direction the overall behavior is clearly not monotonic,
and the magnetic energy-gain has a minimum at a distance $n=4a$.
This is a clear deviation from the RKKY interaction mediated by an
idealized parabolic band.

Finally, we note that the order of magnitude of the interactions in
Figs. \ref{interactions1}, \ref{interactions2} and \ref{gsmagconf}
are in agreement with recent experimental works on Fe atoms deposited
ontop of Pt(111)\cite{Khajetoorians16_Tailoring_chiral_interactions}.

\section{Summary and conclusions}

\label{sec:summary}

We have investigated the indirect spin-spin interactions at the (110)
surface of Pb, mediated by conduction electrons. Our study is motivated
by the luring prospect of engineering spin-spin interactions in nanodevices
with specific functionality at the surface of metals. In particular,
the choice of Pb was motivated by its importance in experiments where
superconductivity and strong spin orbit Rashba interactions (which
emerge due to the lack of inversion symmetry at the surface) are combined.
We have been able to estimate the Rashba parameter as $\alpha_{R}\approx0.97$
eV.{Å}.

In this work, assuming impurity spins $\mathbf{S}_{1}$ and $\mathbf{S}_{2}$,
weakly coupled to the Pb substrate via a generic \textit{s-d} model
(a situation that corresponds to a large class of experimental systems),
we have developed a method which combines realistic \textit{ab initio}
calculations with low-cost computational effort. Using second-order
perturbation theory and realistic DFT calculations for the electronic
band structure of clean Pb(110), we have been able to systematically
obtain the effective spin Hamiltonian between $\mathbf{S}_{1}$ and
$\mathbf{S}_{2}$ at order $J_{K}^{2}$ with no additional assumptions.
Since our method is perturbative, the underlying electronic structure
of clean Pb(110) obtained within \textit{ab initio} is not modified.
Technically, this means that the method, which is suitable for DFT
band-structure calculations based on a periodic lattice, involves
relatively small unit cells. This fact results in a considerable minimization
of computational effort. It is worth mentioning that in general, the
calculation of realistic nanoscale spin interactions through \textit{ab
initio} methods involve a great deal of computational effort (see,
e.g. Ref. \onlinecite{Ebert09_Ab_initio_calculation_of_exchange_coupling}).
Our method allows to systematically track the contribution of the
conduction-electron propagators into the magnetic interaction functions
{[}see Eqs. (\ref{eq:J_RKKY})-(\ref{eq:T_iso}){]}. In addition to
the well-known RKKY interaction, the presence of Rashba spin-orbit
coupling induces a finite DM and tensor matrix interactions, proportional
$\alpha_{R}$ and $\alpha_{R}^{2}$, respectively. In particular the
DM interaction is responsible for non-collinear magnetism and chiral
effects (see \textcolor{black}{Fig. 5}, where we obtain non-collinear
configurations from the minimization of the effective Hamiltonian).
This type of interactions have been recently investigated in relation
to Majorana proposals and skyrmion systems, which is currently investigated
for magnetic storage technology.

The philosophy of our work is reminiscent to those of Imamura \textit{et
al}\cite{Imamura2004}, Zhu \textit{et al}\cite{Zhu11_Electrically_controllable_magnetism_on_TIs}
and Bouaziz \textit{et al}\cite{Bouaziz2017}, where generic indirect
magnetic interactions are obtained directly from the conduction electrons.
However, in contrast to those works we have not assumed any specific
model Hamiltonian for the conduction electrons. In that sense, this
represents an important improvement since it allows to use the knowledge
of realistic band structure calculations. We point out that in many
cases where Rashba spin-orbit coupling is present, there is a tendency
to use phenomenological 2D conduction band models\cite{Imamura2004,Bouaziz2017}.
However, it is known that bulk electrons cannot be neglected and that
they play an important role in, e.g., the Kondo effect\cite{knorr02,Schneider02_Kondo_phase_shift,Limot04}.
A consequence of neglecting bulk electrons is the unrealistic slow
decay of the RKKY and other indirect exchange interactions. In addition,
in certain cases it has been identified that the presence of van Hove
singularities in the idealized 2D band structure produce anomalous
long-ranged interactions\cite{Bouaziz2017}.

Being a perturbative approach based on the second-order expansion
(the ``RKKY approximation''), our method does not take into account
higher-order scattering terms, and therefore is intrinsically limited
to the weak coupling regime $\rho_\text{3D}J_{K}\rightarrow0$. In that
respect, extensions to include higher-order scattering terms have
been proposed in the past\cite{Lloyd72_Multiple_scattering_method_in_condensed_materials,Ebert09_Ab_initio_calculation_of_exchange_coupling,Bouaziz2017}.
%Such implementations are certainly possible within our scheme, as we have full knowledge of the unperturbed electronic propagator $G_{0}$, and they typically involve a self-consistent or a minimization procedure to obtain the groundstate energy configuration for the impurities.
However, it is important to bear in mind that for a magnetic impurity
in the strong-coupling regime, including higher-order scattering terms
might not be enough, and also Kondo correlations, mixed-valence behavior,
charge excitations, and other many-body effects should be addressed
for a proper description. In that respect, Allerdt \textit{et al}\cite{Allerdt15_Kondo_vs_RKKY,Allerdt17_Nonperturbative_effects_exchange_interaction}
have recently considered many-body non-perturbative effects of the
\textit{s-d} exchange on the interaction between quantum impurities
at the surface of metals by implementing the density-matrix renormalization
group (DMRG).

Another important conclusion of our work is the strong anisotropy
of the induced interactions depending on the directionality ($a$
or $b$ directions in Figs. \ref{interactions1} and \ref{interactions2}),
as a result of the symmetry of the Pb(110) surface. This result was
also obtained theoretically\cite{Ebert09_Ab_initio_calculation_of_exchange_coupling}
and experimentally\cite{Zhou2010} in different systems. In addition,
the interaction is non-monotonic with the distance. These two aspects
are in stark contrast with respect to the usually idealized parabolic-band
RKKY.

Finally, we note that the band structure of Pb has been computed for
the normal state, and that the superconducting gap in the spectrum
of quasiparticle excitations of Pb has been ignored. We speculate
that this approximation will not affect our results, as the superconducting
effects should appear at distances of the order of the coherence length
$\xi_{\text{Pb}}\approx80$ nm, which are much larger that the interatomic
distances in our calculations.

In summary, by combining \textit{ab initio} methods with perturbation
theory, we have studied the realistic indirect spin-spin interactions
mediated by conduction electrons in the metallic surface of Pb(110).
We speculate that this approach might be helpful in the design of
weakly-coupled magnetic nanostructures with tailored interactions
in order to obtain specific functionalities.

\acknowledgments{The authors acknowledge financial support from PICT
2017-2081 (ANPCyT-Argentina). A.M.L. acknowledges financial support from PIP 11220150100364 (CONICET - Argentina)
and Relocation Grant RD1158 - 52368 (CONICET - Argentina).}

\bibliographystyle{apsrev4-1}
%\bibliography{references}
%merlin.mbs apsrev4-1.bst 2010-07-25 4.21a (PWD, AO, DPC) hacked
%Control: key (0)
%Control: author (72) initials jnrlst
%Control: editor formatted (1) identically to author
%Control: production of article title (-1) disabled
%Control: page (0) single
%Control: year (1) truncated
%Control: production of eprint (0) enabled
%

\end{document}